\documentclass[twocolumn,preprintnumbers,amsmath,amssymb]{revtex4}

\usepackage{bm} 
\usepackage{graphicx} 

\usepackage{dcolumn} 

\usepackage{natbib}

\newcommand{\mr}{\mathrm} 

\begin{document}
\title{Controlled suppression of superconductivity by the generation of polarized Cooper pairs in spin valve structures}
\author{
M.G.~Flokstra$^{1}$, T.C.~Cunningham$^{1}$, J.~Kim$^{2}$, N. ~Satchell$^{2}$, G. ~Burnell$^{2}$, S.J. ~Bending$^{3}$,  P.J. ~Curran$^{3}$, S.J.~Langridge$^{4}$, C.~Kinane$^{4}$, J.F.K.~Cooper$^{4}$ , N.~Pugach$^{5}$,  M.~Eschrig$^{5}$, and S.L.~Lee$^{1}$
}
\affiliation{ $^{1}$School of Physics and Astronomy, SUPA, University of St.Andrews, KY16 9SS, UK}
\affiliation{$^{2}$School of Physics and Astronomy, University of Leeds, Leeds LS2 9JT,  UK.}
\affiliation{ $^{3}$ Department of Physics, University of Bath, Claverton Down, Bath, BA2 7AY, UK.}
\affiliation{$^{4}$ISIS, Rutherford Appleton Laboratory, Oxfordshire OX11 0QX, UK.}
\affiliation{$^{5}$ SEPnet and Hubbard Theory Consortium, Department of Physics, Royal Holloway, University of London Egham, Surrey, TW20 0EX, UK}

\date{\today}
\begin{abstract}
Transport measurements are presented on thin-film superconducting spin-valve systems, where the controlled non-collinear arrangement of two ferromagnetic Co layers can be used to influence the superconducting state of Nb. We observe a very clear oscillation of the superconducting transition temperature with the relative orientation of the two ferromagnetic layers. Our measurements allow us to distinguish between the competing influences of domain averaging,  stray dipolar fields  and the formation of superconducting spin triplets. Domain averaging is shown to lead to a  weak enhancement of transition temperature for the anti-parallel configuration of exchange fields,  while much larger changes are observed for other configurations, which can be attributed to drainage currents due to spin triplet formation.

\end{abstract}
\pacs{pacs}
\maketitle
\setcounter{figure}{0}

The normally antagonistic ground states of conventional superconductivity and ferromagnetism give rise to a variety of intriguing phenomena when brought in close proximity, a subject that has gained much attention both theoretically \cite{Buzdin-review,Fominov2003,Bergeret2004,Bergeret2005,Matthias,Houzet,Matthias2008,Fominov2010,Wu2012} and experimentally \cite{Kontos2002,Ryazanov,Keizer2006,Salikhov2009,Xia2009,Blamire-2010,Heusler-Zabel,Khaire2010,Aarts2012} over recent years. The underlying proximity effect of singlet Cooper pairs penetrating a ferromagnetic ($F$) layer is non-monotonic in nature, very different from the monotonic decay found for the case of proximity coupling  into a normal ($N$) metal. This unconventional proximity effect leads, for example, to oscillations in the critical temperature ($T_c$) of the superconductor as function of the thickness of the $F$ layer \cite{Radovic91,Jiang95,Jiang97}.

In 2002 the superconducting spin-valve was proposed theoretically \cite{Tagirov,Buzdin99}, comprising a superconducting ($S$) spacer layer separating two $F$ layers. For ideal operation, the super current in the $S$ layer can be controlled by switching the relative orientation of the exchange fields ($H_\mr{ex}$) of the $F$ layers from a parallel (P) to an anti-parallel (AP) alignment. The underlying physical mechanism  involves the interaction of the singlet Cooper pair with both exchange fields, whereby it experiences an additional pair dephasing if the device is in the P-state, due to a potential energy mismatch between the spin up and spin down electron of the penetrated pair, thus lowering $T_c$. Such an effect does not occur in the AP case, since  both electrons find themselves in equivalent bands.  This mechanism can be generalised as a relative enhancement of $T_c$ by domain averaging and has been observed in a variety of experiments \cite{Gu,Rusanov2004,Potenza,Birge2006, Birge2006b}, where, with the exception of ref.~\cite{Rusanov2004}, a {\em pinned} magnetic layer is used to create the AP arrangement.
However, several seemingly anomalous results with  precisely the opposite behaviour have also been reported \cite{Rusanov,Steiner,Stamopolous,Carapella,Flokstra}.  One plausible explanation proposed for these results, in systems where no pinning layer was used, is the dominance of a  suppression of superconductivity by dipolar fields generated by the domains \cite{Rusanov2004}. In experimental work caution therefore needs to be exercised to avoid a dominant contribution from dipolar fields and to be aware that inhomogeneous magnetism (on the length scale of the superconducting coherence length) inherently includes enhancement by domain averaging.

The already rich groundstate in $S/F$ proximity coupled systems becomes even more exotic when non-collinear alignments of the exchange fields are considered. Equal spin triplet pair correlations emerge from the condensate when experiencing inhomogeneous magnetism \cite{Buzdin-review,Bergeret2004,Bergeret2005,Matthias,Houzet,Matthias2008}. Not being an eigenstate of the superconducting condensate, these triplets, unlike singlets,  are not antagonistic to the ferromagnetic ground state and typically penetrate over a much longer distance in $F$ layers (comparable to the case of $N$). This leads to an enhanced drainage of Cooper pairs from the superconductor and thus to a suppression of the superconducting state \cite{Fominov2010}. It was shown theoretically that the density of these spin triplets scales with the magnitude of $H_\mr{ex}$ and one should use strong ferromagnets to observe this suppression.  There are several experiments where the presence of equal spin triplet pairs have been reported \cite{Keizer2006,Blamire-2010,Heusler-Zabel,Khaire2010,Aarts2012} but the generation processes are not fully understood, and  are not always well controlled experimentally.  Experimental data on $S/F$ proximity systems for non-collinear magnetization are vital to better understand these systems and to aid theory towards improved modelling. Some results have been reported on pinned spin-valve type systems \cite{Gu,Potenza,Birge2006,Birge2006b,Zhu2010,Leksin,Tagirov2013},  including angular rotation \cite{Zhu2010,Leksin},  but  to date none have shown an unambiguous enhancement of $T_c$ due to the non-collinearity of the $F$ elements. Most recently experiments on a related  exchange spring system showed results that appear to contradict the predictions of theory in the weak limit \cite{Zhu}.

In this Letter we present transport measurements on Nb/Co based spin-valve systems in which we explore the effect of non-collinear exchange fields on the superconducting state. Our devices were made and characterised  so as to minimize  domain formation and quantify the influence of stray fields,  which enables us to disentangle the observed enhancement of $T_c$ from domain averaging and the suppression by spin triplet drainage.
We observe a {\em large} monotonic increase in the suppression of the superconducting state with increased level of magnetic inhomogeneity, while at collinear angles we recover the established result of the domain averaging effect with an effective $T_c$ shift between the P and AP-state of a few mK.  These results are in strong agreement with theoretical expectations for a  suppression  of $T_c$  with non-collinearity due to the generation of equal spin triplets.

\begin{figure}[tbp]
  \begin{center}
    \includegraphics[width=8cm]{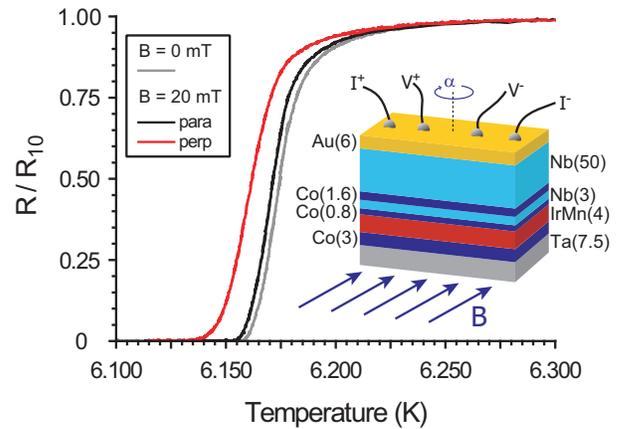}
    \caption{Normalized resistance of the $SFF$ Sample A as a function of temperature for (right) B=0; (middle) B=20 mT parallel to the pinning direction; (left) B=20 mT perpendicular to the pinning direction. Inset: schematic for the $SFF$ structure (Sample A).}
    \label{FIG:RT}
  \end{center}
\end{figure}

We present data on two types of spin-valves, one with the S layer separating the two $F$ layers ($FSF$) and one with the $S$ layer on top ($SFF$) (inset of Fig.~\ref{FIG:RT}). For both architectures, the top $F$ layer is the free layer where the magnetization direction is easily  manipulated by a small external field, while in the bottom $F$ layer the magnetization direction is exchange biased and hence  pinned by an adjacent layer of anti-ferromagnetic IrMn. Samples were prepared by dc magnetron sputtering on Si (100) substrates in a system with a base pressure of 10$^{-8}$ mbar at ambient temperature and in a single vacuum cycle. Growth was undertaken at a typical Ar flow of 24 sccm and pressure of 2-3 $\mu$bar at a substrate-sample distance of approximately 25~mm,  with a typical growth rate of 0.2 nms$^{-1}$. Growth rates for each material were calibrated using fits to the Kiessig fringes in low angle X-ray reflectivity measurements.  All layers were sputtered in the presence of a homogeneous magnetic field at the sample in order to establish the pinning. The full stacking sequence for the $SFF$ spin valve (Sample A) is: 
Au(6)/Nb(50)/Co(1.6)/Nb(3)/Co(0.8)/ IrMn(4)/Co(3)/Ta(7.5)/Si-substrate (inset Fig.\ref{FIG:RT}), with numbers indicating layer thicknesses in nm. The Ta buffer layer is to improve growth quality, the adjacent Co buffer layer is to determine the direction of the pinning for the IrMn and the next Co layer is the actual pinned active layer. The free Co layer is separated from the active pinned Co layer by a thin Nb  decoupling  layer that is non-superconducting. The superconducting Nb layer is next to the free Co layer, and the sample is capped with a thin protective layer to prevent oxidation. The $FSF$  sample (Sample B) has Au(6)/Co(2.4)/Nb(50)/Co(1.2)/IrMn(4)/Co(3)/Ta(7.5) /Si-substrate;   a second $SFF$ structure (Sample C) with Co layers with identical thicknesses to this $FSF$ was also measured. 

Transport measurements on the samples were performed using a standard four-point geometry in a  helium flow cryostat cooled via exchange gas. An external magnet provides a very homogeneous field at the sample. The cryostat  itself is mounted such that it can be rotated around its vertical axis, controlled by a stepper motor. Typical rotation speeds used were about 0.04~rad/sec. Fig.~\ref{FIG:RT} shows a typical transition curve for our devices in zero field (with resistance normalized to the resistance at $T=10$~K) with a $T_c$ of around 6.2~K. Magnetization characterization measurements were performed in a commercial SQUID magnetometer (Quantum Design MPMS-XL) mainly to determine the switching behaviour of the layers.  In addition the stray field for different configurations of the $F$ layers in the $S$ layer  was quantified  using a scanning Hall-probe (SHP) technique, using the microscope described in ref.~\cite{Simon-SHP}.

\begin{figure}[tbp]
  \begin{center}
    \includegraphics[width=8cm]{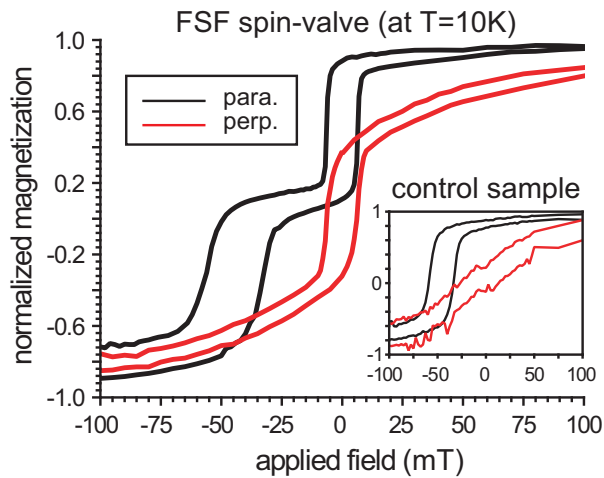}
    \caption{Magnetization measurements on the full $FSF$ spin-valve structure (Sample B)  as function of applied field H$_a$ for (black) H$_a$ parallel to the pinning direction and (red) H$_a$ perpendicular to the pinning direction. Inset: Magnetization measurements on a control sample comprising only a pinned magnetic layer, as a function of applied field H$_a$ for (black) H$_a$ parallel to the pinning direction and (red) H$_a$ perpendicular to the pinning direction.}
    \label{FIG:MH1}
  \end{center}
\end{figure}

To characterize the magnetic switching properties of our spin-valves we first examine a control sample containing only the pinning part of the full device (i.e. omitting the  $S$  and free $F$). Results of SQUID measurements are shown in the inset of Fig.~\ref{FIG:MH1} for the bias (pinning) direction both parallel and perpendicular to the applied field direction of the SQUID. In the parallel case, a clear exchange biased hysteresis curve is obtained with a bias field of around 46~mT, which is associated with the buffer Co layer. The thinner, active pinned layer is much more strongly exchange biased and a slow closing tail is present in the hysteresis curve for negative fields which closes at about -200mT.  A very different response is observed in the perpendicular configuration where no traces of any form of exchange biasing are seen. Note that the tail-part is very similar for both relative orientations.  The perpendicular case can be fitted with a simple Stoner-Wohlfarth model for coherent rotation, including two fixed Zeeman terms to describe the exchange bias fields of the two layers, with one bias around 45~mT and one around 200~mT.

Fig.~\ref{FIG:MH1}  also shows similar magnetization measurements for the full  $FSF$ spin-valve geometry (Sample B). For both orientations the result is like the control sample, with a hysteretic exchange bias of 46~mT and a slow closing tail, but now with an added unbiassed hysteretic part with switching fields of $\pm$6~mT. This corresponds to the response of the free Co layer. The  additional fractional change of the total magnetization is consistent with the fractional Co thickness of the top $F$ layer. The alternative $SFF$ spin-valves (Samples A and C) have almost identical characteristics.

For the rotation transport  measurements we used a fixed external field, typically 16-21~mT, chosen such that it exceeds the switching field of the free $F$ layer but is still well below the exchange bias field of the pinned layers. The sample is rotated with the rotation axis normal to the sample plane.
To further investigate how much stray field is generated  under these conditions we performed SHP measurements, which are sensitive to components of magnetic field perpendicular to the surface of the film. Considering a structure comprising {\em only}  the pinned layer, at a measurement field of $\sim$10~mT, a weak magnetic texture can be observed with a stray field of less than 0.1 mT.  At 77K this texture is found to be {\em totally unchanged} by rotation of the magnetic field between 0$^{o}$ and 90$^{o}$ to the pinning direction. The full $SFF$ structure (Sample A) was also investigate at 10~mT and at a temperature below the $T_c$ of the superconducting layer.   Here once again {\em no variation with angle} of the stray field and magnetic texture was observed for relative orientations of the field to the pinning direction of 0$^{o}$ and 90$^{o}$. We thus conclude that under the conditions in which the transport experiments were undertaken,  there is little contribution from dipolar fields in the superconducting layer, but more importantly that there is a negligible influence on the superconducting state of the dipolar-field contributions as a {\em function of angle}. The results are consistent with a single domain type of rotation of the free layer, while both of the Co layers adjacent to the IrMn layer remains effectively pinned along the bias direction.
\begin{figure}[tbp]
  \begin{center}
    \includegraphics[width=8cm]{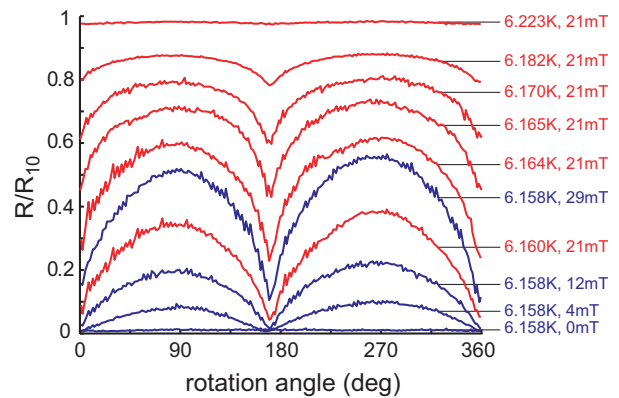}
    \caption{Normalized resistance measurements on the $SFF$ spin-valve structure (Sample A) as function of the angle between the external field and the exchange bias direction, for various temperatures and fields along the transition curve.}
    \label{FIG:RH1}
  \end{center}
\end{figure}
\begin{figure}[h]
  \begin{center}
    \includegraphics[width=8cm]{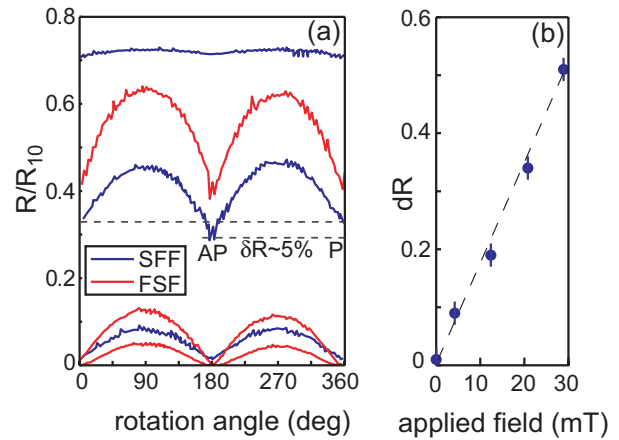}
    \caption{a) Normalized resistance measurements on the $FSF$ (Sample B) and a second $SFF$ spin-valve structure (Sample C) as function of the angle between the external field and the exchange bias direction, for various temperatures along the transition curve. b)  The resistance change $dR$ between $\alpha =0^{o}$ and $\alpha =90^{o}$ induced in the $SFF$ Sample A at 6.158~K by increasing the applied field.}
 \label{FIG:RH2}
  \end{center}
\end{figure}
For an applied field of 21~mT rotating in the plane of the film,  for arbitrary angle the magnetisation of the free layer will  always be parallel to the external applied field,  while the (average) magnetization of the  pinned layers will be coherently tilted by a very small angle away from the pinning direction. At this field one can consider the exchange field in the pinned layer always to be effectively parallel to the pinning direction.
 
Resistance measurements were taken at various positions along the superconducting transition curve as function of the angle $\alpha$ between the external field and the bias direction. The sample was mounted such that the P-state corresponds to $\alpha=0$$^{o}$ and the AP-state to $\alpha=180$$^{o}$. 
All curves presented  are  measured over  a range of 720$^{o}$,  with an average over 2 repeat scans in each direction. 
The voltage noise was found to be dominated by temperature fluctuations which were typically below $\sim$1mK.
Fig.~\ref{FIG:RH1} shows results on the $SFF$ spin-valve (Sample A)  at  different temperatures along the transition curve (the resistance is normalized to the resistance at $T=10$~K). A very clear oscillatory dependence of the resistance as function of $\alpha$ is seen. There are minima near the collinear angles (0$^{o}$ and 180$^{o}$), where the exchange fields are either parallel or anti-parallel to each other, and there are maxima near the perpendicular angles (90$^{o}$ and 270$^{o}$) where those fields are effectively perpendicular.
 The curves approximately follow a $\sqrt{\vert{\sin\alpha}\vert}$ dependence.  
 
Fig.~\ref{FIG:RH2} shows similar measurements on the $FSF$ structure (Sample B) and a second $SFF$ structure (Sample C) with thicker Co layers, 2.4~nm and 1.2~nm for the free and active pinned layer respectively. For both structures the data are qualitatively similar to Fig.~\ref{FIG:RH1} but for the $SFF$ structure with thicker Co layers the oscillations are of smaller amplitude. For temperatures near the steepest parts of the transition curves we can also  
clearly capture the difference between P and AP alignment at the collinear angles. We find a maximum relative resistance difference of about 0.05, which for our typical devices with a transition width of about 50~mK means a corresponding $T_c$ shift of about 2-3~mK (a rather well established result for many spin-valves). This effect is in agreement with theoretical predictions of a slight lowering of $T_c$ for the P state.

The more striking result in the present measurements is the much {\em larger} shifts in $T_c$  observed for all samples due to the {\em non-collinearity} of the magnetic layers.  This can already be clearly seen in the $R(T)$ curves of Fig.~\ref{FIG:RT}.
Theoretically the presence of the non-collinear magnetisation provides a mechanism to increase the conversion of singlet Cooper pairs into the  triplet channel, as has now been observed in a number of experiments involving coherent transport of triplet correlations through a ferromagnetic layer \cite{Keizer2006,Blamire-2010,Heusler-Zabel,Khaire2010,Aarts2012}.  Viewed from the perspective of the singlet superconductor, this represents a 'drainage' current  that partially suppresses the superconducting order parameter and hence lowers $T_c$. Our data are thus in good agreement with these theoretical expectations. Considering the measurements undertaken at 6.158~K for $SFF$ sample A (Fig.~\ref{FIG:RH1}), although there is no angular dependence in zero field, the resistance can be smoothly increased  from zero with both field angle and {\em value of applied field}, indicating a {\em highly tuneable} resistance (Fig.~\ref{FIG:RH2}b). This demonstrates the feasibility of a field-controlled source of equal spin triplets, since it is the drainage to the triplet channel that suppresses the singlet fraction and reduces $T_c$.

We compare our results to a related experiment recently reported on an Nb-Py system, where a Sm-Co exchange spring is used to induce a non-collinear twist in the magnetisation of the Py layer, similar to a Bloch wall \cite{Zhu}. Here the degree of rotation inside the Py layer is controlled by the angle of the applied field to the Sm-Co pinning direction.  In that work a result  is obtained that is superficially opposite to ours. They observe a non-monatonic dependence of applied field with angle that has maxima in the resistivity  at 0$^{o}$ and 180$^{o}$ and minima close to 106 $^{o}$ and 286$^{o}$. As in the present case, these also result from an electronic proximity, but in contrast to our measurements,  the angular dependence appears contradictory to existing theory. We note however that in these experiments the analysis is complicated by the fact that the superconductivity samples only a fraction of the magnetic spiral, and in addition there is a considerable lag between the applied field angle and the total twist angle of the spiral. By contrast in our experiment the coherence length is comparable to the length scale of the magnetic non-collinearity, and at the applied fields measured the field angle is essentially equal to the relative angle of the two exchange fields. It is interesting to note that in the exchange spring experiment a local maximum is observed at an applied field angle of $\theta=180^{o}$, which the authors estimate corresponds to a total {\em non-collinearity} of around $\varphi \sim 90^{o}$, the angle at which we also observe a maximum. This would not of course explain the even larger maximum at $\varphi=0^{o}$ present in their data, but it may be the case that in such a complex magnetic arrangement as an exchange spring that there are  competing influences on the superconducting state. We therefore hope that the very clear results that we present on our much simpler system may help theoretical understanding of these other interesting experiments.

In conclusion we have observed a dependence of the superconducting transition temperature for two different types of Co-Nb spin-valves. For both sample types ($SFF$ and $FSF$)  a large suppression of $T_c$ is  found  when the exchange fields are orthogonal, consistent with the theoretical expectations for the drainage of singlets into the triplet channel when the magnetisation is  non-collinear. This suppression may also be controlled by the magnitude of the applied field. In both structures $T_c$ is a maximum (resistance a minimum) for an AP alignment, with a marginally less pronounced maximum in $T_c$ for the P case, consistent with the  theoretical weak limit result. These results provide a clear and convincing validation of existing theory.  Moreover, since the system is relatively simple, it provides a useful framework in which to understand experimental data in more complex systems where theoretical predictions appear to be contradicted.

\bibliographystyle{unsrt}

\end{document}